\documentclass[12pt]{iopart}
\usepackage{epsfig,graphicx,dcolumn}

\begin{document}

\title{Modeling nonuniversal citation distributions: the role of scientific journals}

\author{Zheng Yao$^{1}$, Xiao-Long Peng$^{1}$, Li-Jie Zhang$^{3,2}$ and Xin-Jian Xu$^{1,2}$}

\address{$^{1}$Department of Mathematics, Shanghai University, Shanghai 200444, People's Republic of China\\
$^{2}$Institute of Systems Science, Shanghai University, Shanghai 200444, People's Republic of China\\
$^{3}$Department of Physics, Shanghai University, Shanghai 200444, People's Republic of China}
\ead{yzhangel380@163.com, xlpeng@shu.edu.cn, lijzhang@shu.edu.cn and xinjxu@shu.edu.cn}

\begin{abstract}
  Whether a scientific paper is cited is related not only to the influence of its author(s) but also to the journal publishing it. Scientists, either proficient or tender, usually submit their most important work to prestigious journals which receives higher citations than the ordinary. How to model the role of scientific journals in citation dynamics is of great importance. In this paper we address this issue through two folds. One is the intrinsic heterogeneity of a paper determined by the impact factor of the journal publishing it. The other is the mechanism of a paper being cited which depends on its citations and prestige. We develop a model for citation networks via an intrinsic nodal weight function and an intuitive ageing mechanism. The node's weight is drawn from the distribution of impact factors of journals and the ageing transition is a function of the citation and the prestige. The node-degree distribution of resulting networks shows nonuniversal scaling: the distribution decays exponentially for small degree and has a power-law tail for large degree, hence the dual behaviour. The higher the impact factor of the journal, the larger the tipping point and the smaller the power exponent that are obtained. With the increase of the journal rank, this phenomenon will fade and evolve to pure power laws.
\end{abstract}

\noindent{\it Keywords}: random graphs, citation networks, scaling

\maketitle

\section{Introduction}

As a treasure trove of quantitative information, the citation dynamics has been studied for a long time \cite{ER90}. A particularly noteworthy contribution was a study in $1965$ by de Solla Price \cite{deSP65}, who proposed the so-called `cumulative advantage' mechanism, that is, a paper
which has been cited many times is more likely to be cited again than others which have been little cited. The cumulative advantage is based on the idea of `the rich get richer' suggested by Yule \cite{YGU25} and Simon \cite{SHA55}, and the criterion now is widely known as the `preferential attachment' in the framework of currently fashionable evolving network models, proposed by Barab\'{a}si and Albert in $1999$ \cite{BA99}. With concepts and methodologies of complex networks, the study of citation patterns has been modeled by networks with nodes representing papers published in scientific journals and edges mimicking citations from papers to others published previously~\cite{HS06,LEA07,WYY08,WH09,ZK10,RC11,GYZ11,GS12,CV12,PRK12,ADR12,ZLL12}.

In citation networks, the probability distribution function appears to involve dual mechanisms with a tipping point between them, that is, a distribution $p(k)$ may decay exponentially for small $k$ and has a power-law tail for large $k$~\cite{RS05,RFC08,CSN09,PWS10}. Here, the degree $k$ represents the number of citations a paper receives and $p(k)$ is the distribution of the relative numbers of such citations. To explain this feature, several mechanisms have been proposed based on the cumulative advantage in different contexts~\cite{JNB03,PPD10,EF11}. For instance, Peterson et al~\cite{PPD10} developed a direct-indirect citing mechanism to fit the citation distribution and analyzed the tipping-point transition. Eom and Fortunato~\cite{EF11} employed the linear preferential attachment with initial attractiveness to reproduce empirical distributions and account for the presence of citation bursts. The cumulative advantage of the node's degree, however, usually weakens and even ignores the hierarchical qualities of papers published in different journals. It is quite possible that a new paper published in a prestigious journal with relatively few current citations will show a potential outbreak on citing, rather than a previously published paper in a common journal with more citations.

On the other hand, the role of time is also important in citation dynamics: old papers are rarely cited while recent papers are usually cited with high frequency. That is, the popularity of papers peaks and then fades. The question of time dependence in the attachment probability of the incoming nodes in a growing network has been addressed in several theoretical models~\cite{DM00,ZWZ03,HS04,HAB07,LR07,CM09,KE02,VA03,TL06,XXJ10}. For instance, Dorogovtsev and Mendes studied the case when the connection probability of the new node with an old one is not only proportional to the degree $k$ but also to a power of its present age $\tau^{-\alpha}$ \cite{DM00}. They found that the network shows scale-free (SF) behaviour only in the region $\alpha < 1$. For $\alpha > 1$, the distribution $P(k)$ is exponential. On the contrary, Klemm and Egu\'{i}luz considered evolving networks based on collective memory of nodes and proposed a degree-dependent deactivation network model \cite{KE02} which transformed the cumulative advantage into the ageing mechanism.

Most previous studies in citation dynamics either focused on a small number of researchers (e.g., high $h$-index scientists) or exploited a database based on one publication society (e.g., papers published in journals of the American Physical Society). This means that they worked on part of the citation network which is not enough to reflect an universal statistical feature on the whole. Of course it is very complicated to analyze citing behaviour of giant papers of the whole citation network or the network of a discipline, such as Physics. It is therefore of great importance to model realistic citation networks with a global view. In this paper we consider the influence of journals on citation dynamics and propose an intuitive mechanism to build the citation network of Physics. We endow each node with a weight for its quality based on the impact factor (IF) of the journal in which it was published and propose a prestige function transformed from the Article Influence Score (AIS) to guide citing. The IF of a journal is the average number of citations received per paper published in that journal during the two preceding years. While the AIS measures the average influence of articles in the journal. Based on these we can represent paper activity not only by the degree of citations but also by the intrinsic weight, which drives the evolution of citation network.

The rest of the paper is organized as follows. In Sec.~\ref{sec2} we describe the indicators from the Journal of Citation Reports, based on which we construct and analyze the model in Sec.~\ref{sec3}. In Sec.~\ref{sec4} we imitate the short-term evolution of the citation network in $2012$ based on the indicators in $2011$. Conclusions are drawn in final Section.

\section{Indicators from the Journal of Citation Reports}\label{sec2}

A complete profile of journals and papers in the Journal of Citation Reports (JCR)~\cite{JCR} includes many indicators for the evaluation of the journals and corresponding papers. We shall use the $5$-Year Impact Factor (IF$5$) instead of the IF as a reasonable weight of a paper published in a journal. The IF$5$ can avoid the influence from outbreak and decline periods of journal citing behaviour. Moreover, its rationality has been validated from different disciplines~\cite{JP09}. Based on the IF$5$, the eigenfactor score takes the heterogeneous importance of other distinct journals which have cited the given journal into consideration. Standardized by the normalized publication volume in the $5$ proceeding years, the eigenfactor score can be transformed into the Article Influence Score (AIS) to intuitively depict the average impact of each paper in a journal. We therefore adopt the AIS as the criterion of paper citing.

We have analyzed almost all the papers in Physics and Astronomy published from $2010$ to $2012$ categorized by the JCR~\cite{JCR} which spread over all the $13$ branches in Physics and $1$ extra branch called Astronomy and Astrophysics. Taking statistics in $2011$ for a sample, there were $174698$ papers published in $652$ journals. The indicators of some journals were not publicated generally for a short founding time or a highly self-cited rate which are unfair to others. After eliminating these defective samples, $168061$ papers in $563$ journals remained, which were assigned with distinct IF$5$s ranged from $0.137$ to $44.436$ and different number of papers published successively. Statistics in the other two years is similar. It should be noticed that the numbers of journals and annual output in physics are steady increasing.

\begin{figure}
  \includegraphics[width=\columnwidth]{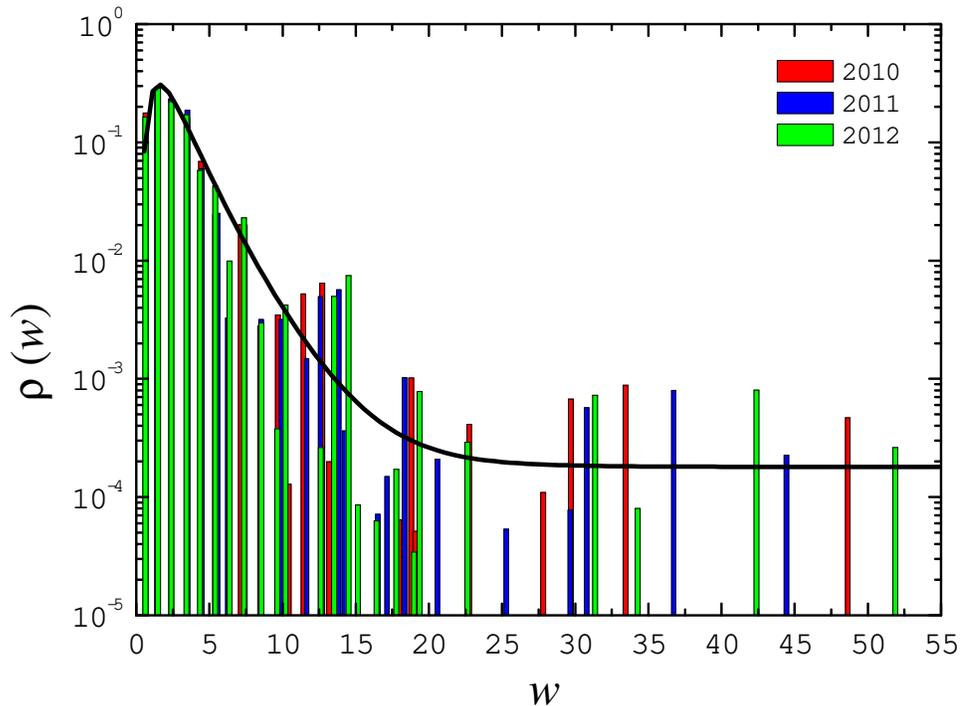}
  \caption{(Color online) The distribution of the normalized publication volume based on the IF$5$ from $2010$ to $2012$. The solid line is the fit in the form of Eq.~(\ref{disp}). The closeness of the $3$-year results implies a stable growth in Physics and Astronomy.}
  \label{fig1}
\end{figure}

Since the IF$5$ of each journal is almost always different, it is rational for one to accomplish the normalized publication volume based on the IF$5$. It should be noticed that some journals with close IF$5$ differ in prestige or academic status in corresponding branch. Also, papers published in the same journal exhibit heterogeneity in citation. As a first step towards modeling paper's heterogeneity, we consider papers published in the same journal having the same intrinsic weight. Although it does not capture accurately and fully the heterogeneity in real data, it will not lead to large influence on the tendency of publication with similar likelihood among them. Thus, we make a reasonable classification for all the journals by distinct intervals of IF$5$s which embodies the specific impact of papers with a certain rank. For simplicity, the journals with IF$5$s between two adjacent integers are aggregated in the same rank. Moreover, the average impact of journals with the same rank, denoted with $w$, can be approximated by the average value of corresponding IF$5$s. Thus, the normalized publication volume of a given rank $\rho(w)$ is accumulated by the number of related articles. The best fit shown in Fig.~\ref{fig1} implies a log-normal distribution,
\begin{equation}
\rho(w)=w_0+\frac{A}{\sqrt{2\pi}\sigma w}\exp\left[-\frac{(\ln w-\mu)^2}{2\sigma^2}\right]
\label{disp}
\end{equation}
with $\mu=0.83(1)$ and $\sigma=0.63(1)$, respectively.

\begin{figure}
  \includegraphics[width=\columnwidth]{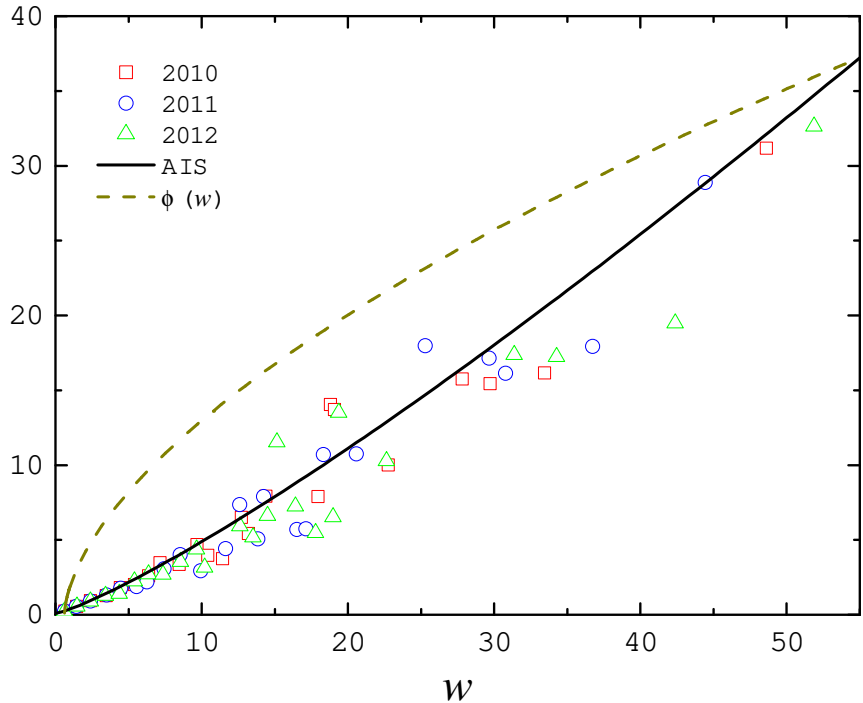}
  \caption{(Color online) Illustration of the AIS and corresponding $\phi(w)$ based on the IF$5$. The original statistics of the AIS is fitted by the solid line, while the dashed line denotes the modulatory function defined by Eq.~(\ref{fh}).}
  \label{fig2}
\end{figure}

\begin{figure}
  \includegraphics[width=\columnwidth]{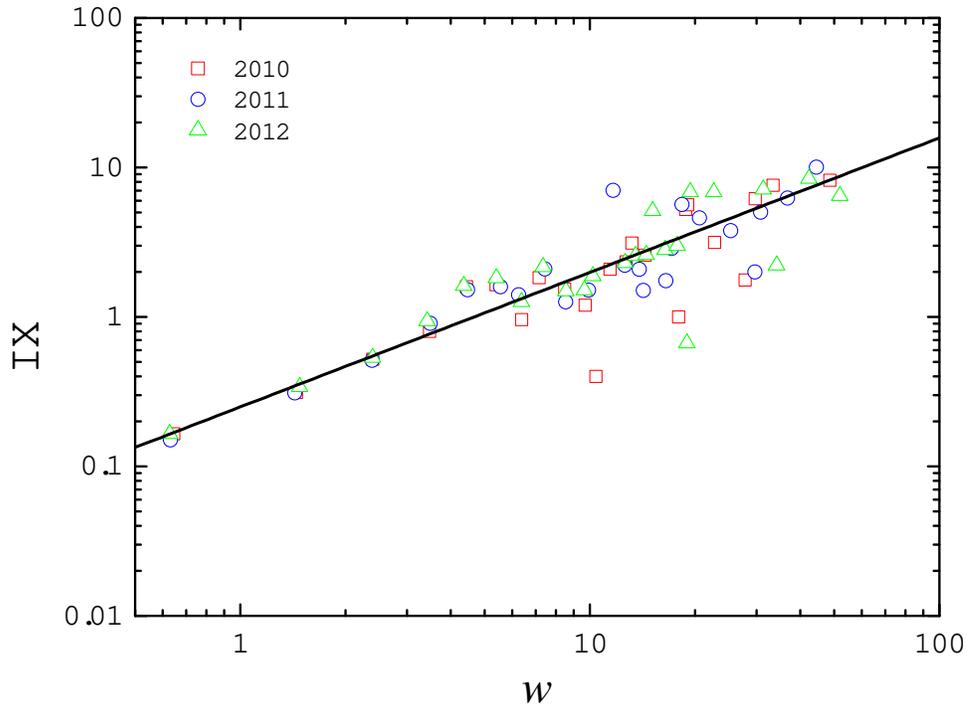}
  \caption{(Color online) Dependence of the IX on the IF$5$ from $2010$ to $2012$. The linear behaviour of the fit implies the power-law scaling.}
  \label{fig3}
\end{figure}

We further investigate the impact of the prestige of journals on citing behavior. As aforesaid, the AIS of a journal weights the impact of each citation of citing journals, and therefore is reasonable for explaining the difference in two journals with close IF. It was found in~\cite{ZLL12} that the AIS and the IF$5$ are positively correlated for most journals. Figure~\ref{fig2} shows the statistics of the AIS for Physics journals with different IF$5$s from the JCR. As one can see, the AIS increases with the IF$5$. The best fit predicts a superlinear growth function. In reality, however, scientists always have a subjective judgment of journals which results in a modulatory effect: for low-AIS papers, the citation increase quickly than the AIS, in other words, scientists are cautious to cite papers published in low-IF journals and a little increase in the IF$5$ will lead to big difference; whereas for high-IF papers, this difference is weak and the speed of the citing increase is lower than that of the AIS. Motivated by this, we propose a new prestige function:
\begin{equation}
\phi(w)=a(w+b)^p,\quad 0<p<1. \label{fh}
\end{equation}

To an extent, the evaluation of papers ($\phi(w)$) stems from scientist's impression on them in recent years. This impression will affect future attention, which can also be reflected by another indicator in the JCR --- Immediacy Index (IX). The IX is defined as the average number of times that a paper, published in a specific year within a specific journal, is cited over the course of the same year, which is a measure of importance and authority of a scientific journal. Figure~\ref{fig3} shows the dependence of the IX on the IF$5$ from $2010$ to $2012$. The best fit suggests a power-law relationship between them. Moreover, the consistency of the $3$-year data informs us that journals with different academic levels steadily maintain its status in Physics research.

\section{Model and analysis}\label{sec3}

In this section we shall use statistical results from the JCR to build the citation network of Physics. Following the idea of Bianconi and Barab\'{a}si~\cite{BB01}, we argue that the evolution of a paper in citation networks depends not only on its topological degree but also on its quality. The topological information, denoted by the indegree $k$, measures citations. While the quality can be characterized by $\rho(w)$. To do that, for each node $i$ a weight $w_i>0$ is assigned, which is the random number drawn from a given probability distribution function $\rho(w)$. Correspondingly, $\phi(w)$ reflects the node's ability to compete for links. With the normalized publication volume $\rho(w)$ and $\phi(w)$ in $2011$, we try to imitate the evolution of the citation network of Physics in $2012$.

The network is initialized with a seed of $n$ isolated nodes which are all active. Then at each time step the dynamics runs as follows.

(i) Add an active node $i$, which connects to $m$ ($m \le n$) nodes randomly chosen from the $n$ active ones.

(ii) Deactivate one (denoted by $j$) of the $n+1$ active nodes with probability
\begin{equation}
\pi (k_j, w_j) =\sigma [(k_j+c)\phi(w_j)]^{-1}, \label{deactprob}
\end{equation}
where $\sigma=\{\sum\limits_{j\in \Lambda}[(k_j+c)\phi(w_j)]^{-1}\}^{-1}$ is the normalization factor and $c$ is a constant bias. The summation runs over the set $\Lambda$ of the $n+1$ active nodes. Eq.~(\ref{deactprob}) implies that a paper with high citations or prestige has a small possibility to be forgotten. During evolution, a node might receive edges while it is active, and once inactive it will not receive edges any more. Moreover, the self-cited phenomenon hardly happens due to the uncorrelated attachment mechanism.

Denoting with $a(k,w,t)$ the number of active nodes with indegree $k$ and weight $w$ at time $t$, one can write out the rate equation for network evolution:
\begin{eqnarray}
\frac{\partial a(k,w,t)}{\partial t} & = \frac{m}{n}\left[1-\pi(k-1,w)\right]a(k-1,w,t)-\frac{m}{n}a(k,w,t) \nonumber \\
& - \left(1-\frac{m}{n}\right)\pi(k,w)a(k,w,t)+\left[1-\pi(0,w)\right]\rho(w)\delta_{k,0}. \label{mdiff}
\end{eqnarray}
On the right-hand side of Eq.~(\ref{mdiff}), the first term represents an active node with indegree $k-1$ and weight $w$ at time $t$ is connected by the newcomer and not deactivated at current time, the second and third terms account for the reduction of $a(k,w,t)$, and the last term is a complement for the boundary condition that the newcomer with weight $w$ is not deactivated at current time. Imposing the stationarity condition $\partial a(k,w,t)/\partial t=0$ yields
\begin{eqnarray}
a(k,w) & = a(0,w)\frac{k+c}{c}\exp \{ \sum\limits_{i=1}^k\ln\frac{m[(i+c-1)\phi(w)-\sigma]}{[m(i+c)f(w)+\sigma(n-m)]} \}, \label{diffeq1}\\
a(0,w) & = \frac{n\rho(w)[c\phi(w)-\sigma]}{\sigma n+m[c\phi(w)-\sigma]}. \label{diffeq2}
\end{eqnarray}
Employing Stirling approximation, the solution can be simplified to
\begin{eqnarray}
a(k,w) & \approx a(0,w)\left[c-\frac{\sigma}{\phi(w)}\right]^{-[c-\frac{\sigma}{\phi(w)}]}\left[1+c+\frac{\sigma(n-m)}{m\phi(w)}\right]^{1+c+\frac{\sigma(n-m)}{m\phi(w)}} \nonumber \\
& \times \frac{k+c}{c}\frac{\left[k+c-1-\frac{\sigma}{\phi(w)}\right]^{k+c-1-\frac{\sigma}{\phi(w)}}}{\left[k+c+\frac{\sigma(n-m)}{m\phi(w)}\right]^{-[k+c+\frac{\sigma(n-m)}{m\phi(w)}]}}. \label{diffeq3}
\end{eqnarray}
Denoting with $D(k,w,t)$ the number of inactive nodes with indegree $k$ and weight $w$ at time $t$, it obviously satisfies
\begin{equation}
\sum\limits_w\sum\limits_k[D(k,w,t+1)-D(k,w,t)]=1,
\end{equation}
which means only one inactive node is generated at each time step so that $D(k,w,t)$ is increased by one at next time. Thus, we obtain the master equation for $D(k,w,t)$:
\begin{equation}
D(k,w,t+1)=D(k,w,t)+{\rm Prob} \{\Delta D(k,w,t)=1\}.
\end{equation}
Since $D(k,w,t)$ grows linearly with $t$, we introduce the function $d(k,w)$ such that
\begin{equation}
D(k,w,t)=td(k,w),
\end{equation}
where $d(k,w)$ is the stationary probability of inactive nodes with indegree $k$ and weight $w$. In case that the total number $N$ of nodes in the network is far more than the number $n$ of active nodes, the degree distribution $p(k,w)$ can be approximated by considering inactive nodes only, i.e.,
\begin{eqnarray}
p(k,w) & \approx d(k,w) & = -\frac{m}{n}\frac{\partial a(k,w)}{\partial k},\quad k\geq 1, \label{rel2} \\
p(0,w) & \approx d(0,w) & = \rho(w)-\frac{m}{n}a(0,w). \label{rel3}
\end{eqnarray}
Substituting Eqs.~(\ref{diffeq2}) and (\ref{diffeq3}) into Eqs.~(\ref{rel2}) and (\ref{rel3}) respectively, one obtains
\begin{eqnarray}
p(k,w) & = \frac{\sigma m\rho(w)[c\phi(w)-\sigma]}{\{\sigma n+m[c\phi(w)-\sigma]\}c}
\frac{\left[1+c+\frac{\sigma(n-m)}{m\phi(w)}\right]^{1+c+\frac{\sigma(n-m)}{m\phi(w)}}}{\left[c-\frac{\sigma}{\phi(w)}\right]^{c-\frac{\sigma}{\phi(w)}}} \nonumber \\
& \times \left[\frac{k+c-1-\frac{\sigma}{\phi(w)}}{k+c+\frac{\sigma(n-m)}{m\phi(w)}}\right]^{k+c-\frac{\sigma}{\phi(w)}} \ln \{ \frac{1}{e}\left[\frac{k+c+\frac{\sigma(n-m)}{m\phi(w)}}{k+c-1-\frac{\sigma}{\phi(w)}}\right]^{k+c} \} \nonumber \\
& \times \left[k+c-1-\frac{\sigma}{\phi(w)}\right]^{-1}\left[k+c+\frac{\sigma(n-m)}{m\phi(w)}\right]^{-\frac{\sigma n}{m\phi(w)}}, \label{pk1} \\
p(0,w) & = \frac{\sigma n \rho(w)}{\sigma(n-m)+mc\phi(w)}. \label{pk2}
\end{eqnarray}

\section{Short-term evolution}\label{sec4}

To imitate the evolution of the citation network of Physics in $2012$, one needs qualitative information of model parameters according to the JCR in $2011$. For simplicity, we let $m=1$ and $n=14$, which assumes one popular article within each physical branch initially. $c$, related to the tipping point between the exponential and the power law, is set $10$ just as well. Considering the magnitude of the annual output from $2010$ to $2012$, we let $N=10^5$. The distribution $\rho(w)$ has already been given by Eq.~(\ref{disp}). As to $\phi(w)$, since there is a one-to-one correspondence between the IX and $w$ (see Fig.~\ref{fig3}), one can obtain the realistic $\phi(w)$ from the statistics. In fact, the distribution of the IX based on different journals just matches the conditional expectation $\langle k|w \rangle$ in our model, which is the distribution of the average indegree of nodes with distinct weights. Therefore, by analyzing $\langle k|w \rangle$ we can obtain exact $\phi(w)$.

Before computing $\langle k|w \rangle$, one needs another operation. From the JCR, the average IX of all the papers in Physics, counted as $0.62$ in $2010$, $0.64$ in $2011$ and $0.71$ in $2012$, respectively, grows steadily in recent years, which implies that the frontier of papers published by various journals in Physics and Astronomy persists with high quality. Since our aim is to predict what will happen in the citation network of Physics in $2012$, we only take papers published in $2012$ into consideration and the citations from new papers to those published before $2012$ can be neglected, in other words, the outdegree $m$ of each new node should be small, set $1$ in this work. we therefore enlarge uniformly the fitting curve in Fig.~\ref{fig3} with a ratio of $1/0.71$ to eliminate the scale error.

\begin{figure}
  \includegraphics[width=\columnwidth]{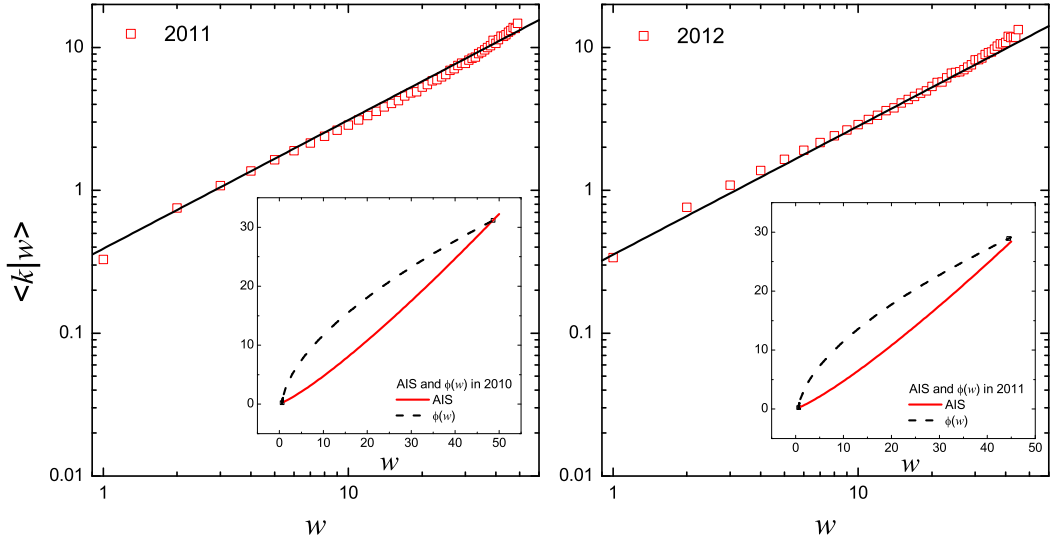}
  \caption{(Color online) Comparison of $\langle k|w \rangle$ with the IX. The solid lines are the enlargement of the fits from Fig.~\ref{fig3} by the ratios $1/0.64$ for $2011$ (a) and $1/0.71$ for $2012$ (b), respectively. The insets of the both plots respectively show the original AIS and the predicted $\phi(w)$ for $2010$ and $2011$.}
  \label{fig4}
\end{figure}

From Eqs.~(\ref{pk1}) and (\ref{pk2}), one easily obtains
\begin{eqnarray}
p(k|w) & = \frac{p(k,w)}{\rho(w)},\quad k\geq 1, \label{pk3} \\
p(0|w) & = \frac{\sigma n}{\sigma(n-m)+mc\phi(w)}, \label{pk4}
\end{eqnarray}
which yields
\begin{equation}
\langle k|w \rangle = \sum\limits_{k=1}^{k_{max}(w)}\frac{kp(k,w)}{\rho(w)}. \label{id}
\end{equation}
It is hard to estimate the maximal indegree with weight $w$, so we always take a large value for each case, which may bring about the error. Also, Eq.~(\ref{id}), which is too complicated to be simplified, is restricted to numerical calculations. In Fig.~\ref{fig4}, we find a consistence between the simulation result of $\langle k|w \rangle$ and the fitting line in Fig.~\ref{fig3} when the exponent $p$ in Eq.~(\ref{fh}) in $2011$ is $0.6(5)$. The good agreement of the numerical simulations with real data in the JCR demonstrates the conjecture on $\phi(w)$. On this promise, we can make a short-term prediction of the citation network in Physics and Astronomy based on certain $\rho(w)$ and $\phi(w)$.

Different from work by Peterson et al. \cite{PPD10}, the present model is more reasonable on the cross reference among distinct journals. We therefore focus on the conditional probability $p(k|w)$ which aggregates the citation statistics among journals with a certain rank, instead of the total probability $p(k)$. According to Eqs.~(\ref{pk1}) and (\ref{pk3}), we obtain
\begin{eqnarray}
p(k|w) & \sim \left[\frac{k+c-1-\frac{\sigma}{\phi(w)}}{k+c+\frac{\sigma (n-m)}{m\phi(w)}}\right]^{k+c-\frac{\sigma}{\phi(w)}}\ln \{ \frac{1}{e}\left[\frac{k+c+\frac{\sigma (n-m)}{m\phi(w)}}{k+c-1-\frac{\sigma}{\phi(w)}}\right]^{k+c} \} \nonumber \\
& \times \left[k+c-1-\frac{\sigma}{\phi(w)}\right]^{-1}\left[k+c+\frac{\sigma(n-m)}{m\phi(w)}\right]^{-\frac{\sigma n}{m\phi(w)}}. \label{pk5}
\end{eqnarray}
In the small-$k$ region ($k\ll c-\displaystyle\frac{\sigma}{\phi(w)}$), we have
\begin{equation}
p(k|w) \sim \{ \frac{mc\phi(w)+\sigma(n-m)}{mc\phi(w)-m[\sigma+\phi(w)]} \}^{-[k+c-\frac{\sigma}{\phi(w)}]}.
\end{equation}
Whereas for the large-$k$ region ($k\gg c-\displaystyle\frac{\sigma}{\phi(w)}$), Eq.~(\ref{pk5}) can be simplified to
\begin{equation}
p(k|w) \sim \left[k+c-\frac{\sigma}{\phi(w)}\right]^{-[1+\frac{\sigma n}{m\phi(w)}]}.
\end{equation}
Between them, there is a tipping point:
\begin{equation}
k^{*}=c-\frac{\sigma}{\phi(w)}, \label{tip}
\end{equation}
which implies that the more prestigious a journals is, the more intensely the papers published by it compete for links. Papers initialized with a small degree $k$ will receive random attention unless they have got abundant citations. On the other hand, an important result about the nonuniversal power-law scaling is determined by
\begin{equation}
\gamma(w)=1+\frac{\sigma n}{m\phi(w)}. \label{gamma}
\end{equation}

\begin{figure}
  \includegraphics[width=\columnwidth]{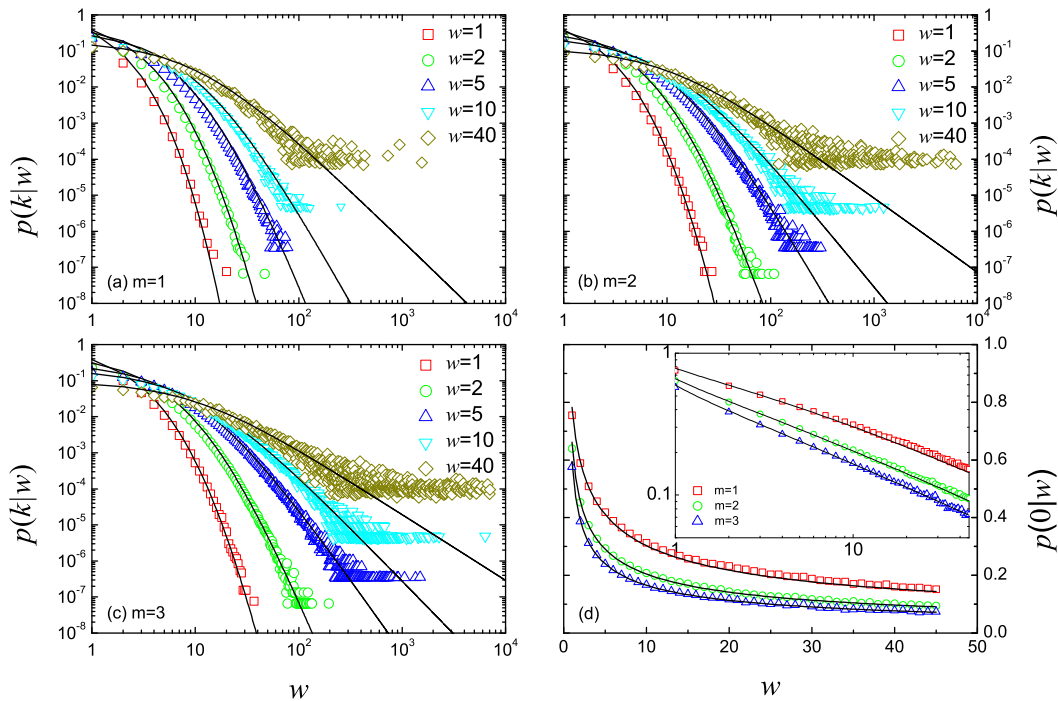}
  \caption{(Color online) Simulation results of the citation distributions of journals with distinct ranks. Three values of $m=1$, $2$, $3$ have been set to carry out the short-term prediction. With the assumption of a balanced resource allocation, the link density $m/n$ remains unchanged. As one can see, there are exponential, exponentially truncated power-law and power-law distributions in the large-$k$ tail. All the solid lines are the predictions of Eqs.~(\ref{pk3}) and (\ref{pk4}), respectively.} \label{fig5}
\end{figure}

\begin{figure}
 \includegraphics[width=\columnwidth]{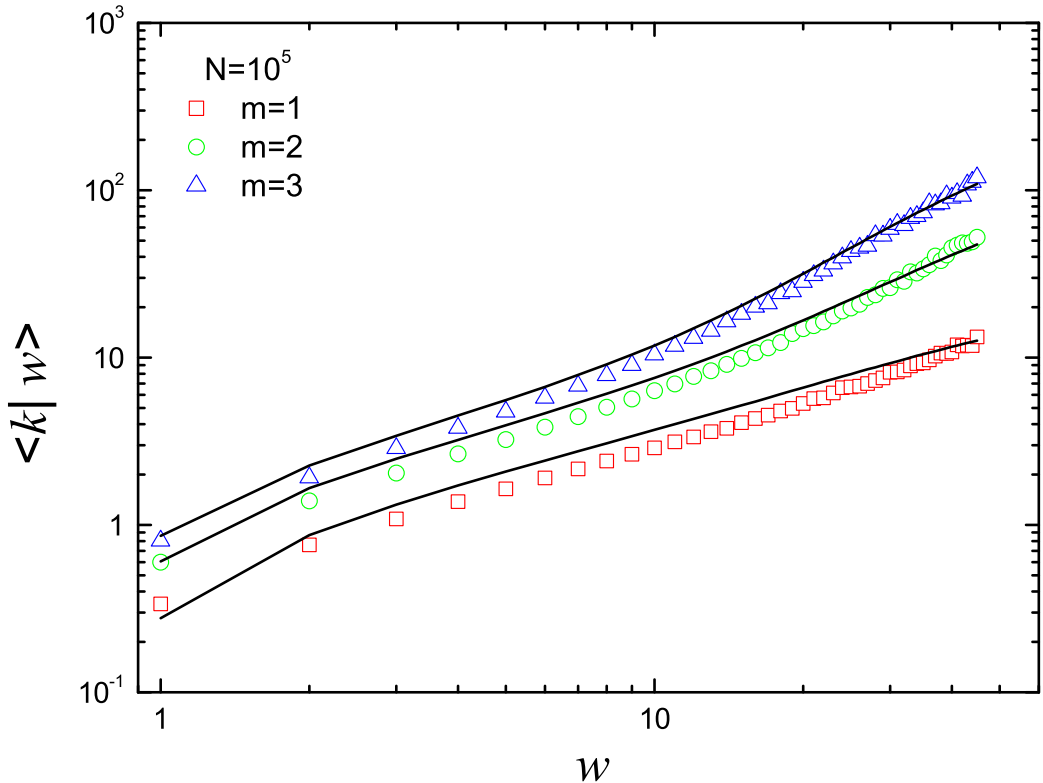}
 \caption{(Color online) Plots of $\langle k|w \rangle$ as a function of $w$. The solid lines correspond to the predictions of Eq.~(\ref{id}), the error of which results in the difference between the analytical solutions and the numerical ones.}
 \label{fig6}
\end{figure}

From Fig.~\ref{fig5}, we find that distributions of low-IF$5$ journals are exponentially truncated power-law, and even exponential in the large-$k$ region. The smaller $w$, the more exponentially they decay. Moreover, the tipping point $k^{*}$ is not sensitive to $w$. However, there are hardly any highly-cited papers published by low-IF$5$ journals, which is reflected intuitively from the monotonous relationship between the IX and the IF$5$ (see Fig.~\ref{fig4}). It is very difficult for these journals to accumulate high citations in a short period. With so little attractiveness, even certain papers with relatively high citations can not show a great advantage over $k^{*}$, which results in the exponential distribution of $p(k|w)$. That is, the behaviour of the citation distribution in the large-$k$ region will transit from the exponential to the exponentially truncated power law, and finally the power law, with the increase of the IF$5$. This transition also emerges in network evolution. For simplicity, we assume the attractiveness from new papers to old ones to be homogeneous in a short period such as $1$, $2$ or $3$ years. Based on this, we can roughly investigate a short-term evolution of the citation network in Physics. From Figs.~\ref{fig5}(a) to \ref{fig5}(c), we find that the exponential behaviour in the large-$k$ region is gradually replaced by the power law with the increase of $m$. This transition of network structure results from the intrinsic scientific evolution. Both papers and their references have been shown to increase \cite{PRK12}, which will give rise to network connectivity. Denoting with $2N=2\times 10^5$ the $2$-year output, a continuous inequality is formed:
\begin{equation}
\langle k|w \rangle_{2N,m=2}>\langle k|w \rangle_{N,m=2}>\langle k|w \rangle_{N,m=1}. \label{khNm}
\end{equation}
Given the weight $w$, the increase of papers only contributes to the maximal indegree of nodes since older papers can accumulate their advantage, which gives rise to the first inequality. While the second inequality can only be numerically verified because of the uncertain change in $\sigma$ with the increase of the outdegree. Figure~\ref{fig6} shows the rising of $\langle k|w \rangle$ with $m$. Meanwhile, no matter how $k^{*}$ changes, it will not exceed $c$ according to Eq.~(\ref{tip}). Therefore, excellent papers will attract more attention as time goes by, which might show a great advantage over $k^{*}$ at a certain period. As a consequence, the citation network evolves from the exponential to the heterogeneous. Another interesting feature is the power-law behavior of the non-cited papers, as shown in Fig~\ref{fig5}(d), the larger IF$5$ of the journal, the fewer non-cited papers were published by them, which characterizes the impact of distinct journals from another perspective.

\section{Conclusion}

To summarize, we have presented a kind of driving force of scientific progress dually based on the selection of publication and the evaluation of citation respectively according to journals with distinct IF$5$s and articles with different AISs to investigate the short-term evolution of the citation network of Physics with real data from the JCR. We defined the IF$5$ as the intrinsic weight of a journal, based on which the normalized publication volume $\rho(w)$ and a reasonable evaluation function $\phi(w)$ have been obtained. The former describes paper's hierarchy while the later characterizes the influence or impression from collective memory. The growth dynamics of the proposed network model combines the addition of new active nodes randomly connected to existing active ones and the transition of active nodes to the inactive. The deactivation probability of a node is inversely proportional not only to its degree $k$ (citations) but also to the evaluation function $\phi(w)$ (prestige).

We have employed the model to imitate the evolution of the citation network of Physics in short period and observed two kinds of nonuniversal scaling phenomena. The citation distribution of journals with same rank decays exponentially for small $k$. In the large-$k$ region, however, we obtained continuous transitions from the exponential to the exponentially truncated power law, and finally the power law, with the increase of the IF$5$. Meanwhile, the increase of $m$ also leads to the same process. Moreover, we found that the behaviour of the power law also depends on the IF$5$. The higher IF$5$ of a journal, the larger tipping point and the smaller power exponent were obtained. All these results should be some common regularities in other disciplines.

The heterogeneity of papers is of great importance in citation dynamics. The present model addressed this issue from the perspective of scientific journals, that is, the quality of a paper is defined by the IF$5$ of the journal in which it was published. To distinguish papers published in two different journals with close IF$5$, we have introduced the AIS to weigh different impacts of papers by the definition of the prestige function. However, the heterogeneity of papers published in the same journal was ignored in this work. Therefore, the present model is a first step toward modeling the heterogeneity of papers in citation networks. The large heterogeneity among different fields, authors and collaborations will draw future attention.

\section*{Acknowledgments}

This work was jointly supported by NSFC/11331009, SHMEC/13YZ007 and STCSM/13ZR1416800.

\end{document}